\documentclass[reprint,superscriptaddress,amsmath,amssymb,aps,prl,floatfix]{revtex4-2}
\usepackage{graphicx}
\usepackage{dcolumn}
\usepackage{bm}
\usepackage{hyperref}
\usepackage{xcolor}
\usepackage{float}
\usepackage{bigints}
\usepackage{nameref}

\definecolor{change}{rgb}{1., 0., 0.}

\definecolor{darkred}{HTML}{8B1E2D}

\definecolor{darkgreen}{HTML}{00660C}

\newcommand{\cavr}[1]{\langle\!\langle #1\rangle\!\rangle}

\begin{document}

\title{Hints for a QCD critical end-point from simulations on a finite lattice}
\author{Szabolcs Bors\'anyi}
\affiliation{Department of Physics, Wuppertal University, Gaussstr.  20, D-42119, Wuppertal, Germany}

\author{Zolt\'an Fodor}
\affiliation{Pennsylvania State University, Department of Physics, State College, PA 16801, USA}
\affiliation{Pennsylvania State University, Institute for Computational and Data Sciences, University Park, PA 16802, USA}
\affiliation{Department of Physics, Wuppertal University, Gaussstr.  20, D-42119, Wuppertal, Germany}
\affiliation{Institute  for Theoretical Physics, ELTE E\"otv\"os Lor\' and University, P\'azm\'any P. s\'et\'any 1/A, H-1117 Budapest, Hungary}
\affiliation{J\"ulich Supercomputing Centre, Forschungszentrum J\"ulich, D-52425 J\"ulich, Germany}

\author{Jana N. Guenther}
\affiliation{Department of Physics, Wuppertal University, Gaussstr.  20, D-42119, Wuppertal, Germany}

\author{Piyush Kumar}
\affiliation{Department of Physics, Wuppertal University, Gaussstr.  20, D-42119, Wuppertal, Germany}

\author{Paolo Parotto}
\affiliation{Dipartimento di Fisica, Universit\`a di Torino and INFN Torino, Via P. Giuria 1, I-10125 Torino, Italy}

\author{Attila P\'asztor}
\affiliation{Institute  for Theoretical Physics, ELTE E\"otv\"os Lor\' and University, P\'azm\'any P. s\'et\'any 1/A, H-1117 Budapest, Hungary}
\affiliation{MTA-ELTE Lendület "Momentum" Strongly Interacting Matter Research Group, Budapest, Hungary}

\author{D\'avid Peszny\'ak}
\affiliation{Department of Physics, Wuppertal University, Gaussstr.  20, D-42119, Wuppertal, Germany}

\author{Chik Him Wong}
\affiliation{Department of Physics, Wuppertal University, Gaussstr.  20, D-42119, Wuppertal, Germany}

\date{\today}

\begin{abstract}
A novel approach is introduced to lattice QCD to study the finite density phase structure. In particular, we calculate 
the constraint effective potential of the Polyakov loop based on 
the density-of-states formalism at finite baryo-chemical potential. The superior convergence of this resummed expansion opens unexplored regions of the phase diagram. 
It allows extrapolations through a transition. We simulate
three low temperatures with the lattice spacing of $a=0.2$~fm
and compute the NNLO potential. The behavior
at $T=120$~MeV is consistent with a crossover, at 80~MeV we see
signs typical for a first order transition. At $T=96$~MeV we see a flat potential as expected near a critical point.
\end{abstract}

\maketitle

\paragraph{Introduction}

An important open problem in high energy physics is the determination of the phase diagram of QCD in the
temperature ($T$) - baryochemical potential ($\mu_B$) plane. From a theoretical perspective, lattice QCD simulations at $\mu_B=0$ show a smooth chiral crossover around $T=158$~MeV \cite{Aoki:2006we,Borsanyi:2020fev}. The static quark entropy (defined via the temperature dependence of the Polyakov loop) also shows a peak in the same temperature range~\cite{Bazavov:2016uvm, Borsanyi:2025lim}.
Lattice simulations have also traced how this crossover temperature changes with a small real $\mu_B$~\cite{Borsanyi:2020fev, HotQCD:2018pds, Bonati:2018nut, Borsanyi:2024xrx}.
Whether, by increasing $\mu_B$ further, this crossover ends at a critical point and gives way to a first order transition remains an open question. Functional calculations have yielded estimates of where such a critical point could lie \cite{Fu:2019hdw,Gunkel:2021oya,Wang:2026xwa,Fu:2026qnl,Gao:2026pfr}.

The sign problem makes direct lattice simulations at real $\mu_B$ difficult. 
One way to reach finite density is to expand thermodynamic observables
around $\mu_B=0$~\cite{Allton:2002zi}. However, studying phase transitions with Taylor expansions has difficulties. At a second order transition, convergence is limited by the critical point. At a first-order transition, the expansion (at infinite volume) can converge to the wrong value, as it only provides the analytic continuation of the energy density for one state, even in the region where it is only metastable.

In a finite volume, both phases will be present, and the
distribution of an observable that distinguishes 
can be bimodal.
The relative weights of the two peaks show how the balance between
the phases changes across the transition region. The presence of such a 
distribution leads to an overlap problem.

Density-of-states methods can handle such overlap problems. One introduces a constraint in an observable, forcing the simulation to visit regions of the configuration space it otherwise wouldn't, and then integrates over the constraint.
In finite density lattice QCD, constraining the plaquette has been tried in Refs.~\cite{Fodor:2007vv,Ejiri:2007ga}. In this Letter we 
constrain the real part of the Polyakov loop, which is
sensitive to 
the QCD crossover.
We then use derivatives of the constrained
partition function to expand the potential in $\frac{\mu_B}{T}$ at fixed loop
value. Integrating the Boltzmann weight over the loop gives the full
partition function. A smooth change in the potential can 
change which local minimum dominates the integral. 

In other words, we are expanding the constraint effective effective potential of the Polyakov loop before integration over possible loop values. This amounts to a resummation
of the pressure. The ordinary Taylor coefficients are recovered
up to the order retained in the potential, but any finite truncation in the potential
generates contributions in the pressure to infinite orders in $\mu_B$. 
A similar resummation was tested (and found to have excellent convergence properties) in Ref.~\cite{Mukherjee:2021tyg}, where the mean-field potential of the $\sigma$ field was expanded in a quark-meson model, before minimization.

We perform a sixth order resummation in a finite volume, physical masses and a lattice spacing of
$a=0.2$~fm. At $T=80$~MeV a double well appears
at all three truncation orders when each potential is evaluated at
its own transition point estimate. At $T=96$~MeV we find a broad
minimum, while at $T=120$~MeV we find a single parabolic well. This is consistent with a first-order line
ending at a critical point. The chemical potentials associated with
these features shift as higher orders are included. 
Full control of truncation, volume, and cut-off effects is left for future work.

\begin{figure*}[t]
\centering
\includegraphics[width=0.99\textwidth]{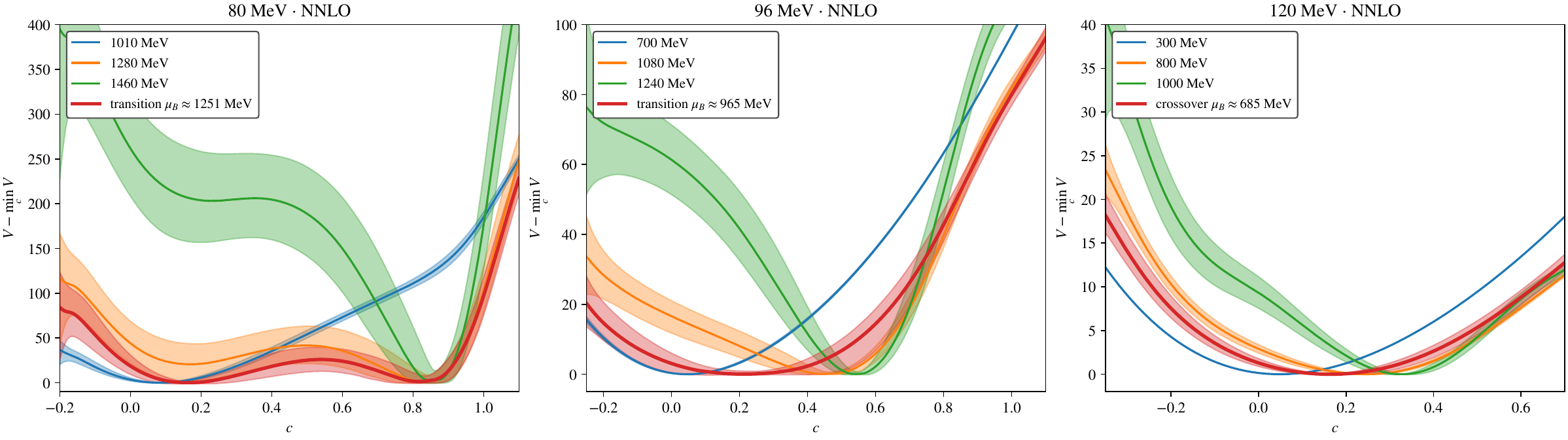}
\caption{\label{fig:potential_scans}
NNLO order computation of the Polyakov loop effective potential
at three temperatures in our finite lattice spacing study.
Data at the lowest temperature exhibits a double-well potential,
typical for a first order temperature, while at $96$~MeV we observe
flat potential. The potential at the highest temperature is consistent
with a cross-over. The transition point in $\mu_B$ is given
for all three cases in the legend. We emphasize, that 
additional orders are necessary for this expansion's convergence. The chemical potential for the
red curves were tuned such that the two
minima are of equal height (left panel),
or the curvature is the smallest possible in the
minimum (middle and right panels.)
}
\end{figure*}

\paragraph{Formalism}

The central object in this approach is the constraint effective potential $V(\phi)$ of the order parameter. In this work we choose $\phi=(\ell+\bar\ell)/2$ where $\ell$ and $\bar\ell$ are the fundamental and the conjugated Polyakov loops, respectively.

$V(\phi)$ is typically explored only near its true minimum in a lattice simulation. At the onset of a transition at high $\mu_B$ previously un-visited parts of  the configuration space will become dominant, e.g. with a significantly larger order parameter value. Already a small change in the $V(\phi)$ function can account for such a transition.

Ideally, one would directly simulate the expectation value $\langle \phi\rangle$ as a function of a control parameter at high $\mu_B$
and observe a strong crossover or a true transition. Since
such direct data taking is hindered by the sign problem one resorts to
extrapolations. The traditional Taylor expansion of the order parameter will 
fail by construction to deliver beyond the point of the transition, or, beyond the
magnitude of the closest Yang-Lee singularity on the complex plane of the
chemical potential. Thus, our focus will be in this work to apply the Taylor
extrapolation on the effective potential itself. This means that the effective potential will be extrapolated at fixed order parameter value ($\phi)$ as
\begin{eqnarray}
V(\phi,\mu_B) = V_0(\phi) 
- \chi_2(\phi) \frac{\hat\mu_B^2}{2!}
- \chi_4(\phi) \frac{\hat\mu_B^4}{4!} + \dots
\label{eq:Vexpansion}
\end{eqnarray}
with $\hat\mu_B=\mu_B/T$. This expansion is understood at fixed temperature $T$, selected below the known crossover temperature of 158~MeV at $\mu_B=0$ \cite{Borsanyi:2020fev}.
This effective potential is directly related to the full partition sum
\begin{eqnarray}
Z(\mu_B) = \int d\phi\, e^{-(LT)^3\,V(\phi,\mu_B)}
\label{eq:ZwithV}
\end{eqnarray}
where $L^3$ is the spatial volume, thus, defining $V$ as a dimensionless
quantity. We only consider order parameters that are
symmetric under charge conjugation, in this case $V(\phi,\mu_B)$ and its
coefficients are real valued, and the odd $\chi_n(\phi)$ coefficients vanish.
From Eq.~(\ref{eq:ZwithV}) the expectation value of the
order parameter immediately follows
\begin{eqnarray}
\langle\phi\rangle = 
\frac{1}{Z(\mu_B)} \,
 \int d\phi\,\phi\, e^{-(LT)^3\,V(\phi,\mu_B)}\,.
\end{eqnarray}
Away from a true transition, this integral is dominated by the vicinity
of the actual minimum of the potential ($\phi_0$), and we will find
$\langle\phi\rangle\to\phi_0$
in the thermodynamic limit $(LT\to\infty)$.
Unlike the effective potential defined
via the Legendre transform the constraint
effective potential $V(\phi)$ can be nonconvex,
similar to potential in the mean field approximation.

Inserting the expansion Eq.~(\ref{eq:Vexpansion}) into Eq.~(\ref{eq:ZwithV})
fully defines the proposed expansion scheme, from this all densities and
susceptibilities can be computed at finite $\mu_B$.
Consider for example the baryon density 
\begin{eqnarray}
\frac{\langle n_B\rangle}{T^3} &=&
\frac{1}{(LT)^3}
\frac{\partial \log\,Z(\mu_B)}{\partial \hat\mu_B} 
\label{eq:nB}
\\
&=&
\frac{1}{Z(\mu_B)}\, 
 \int d\phi\,\left(\frac{\partial (-V)}{\partial \mu_B}\right)\, e^{-(LT)^3\,V(\phi,\mu_B)}\,.\nonumber
\end{eqnarray}
The leading order expansion of $V$ is written up to the term
$\chi_2\hat\mu_B^2/2$ in Eq.~(\ref{eq:Vexpansion}), at this order
we find for $\langle n_B\rangle$
\begin{eqnarray}
\frac{\langle n_B\rangle}{T^3} &=&
\hat\mu_B\cdot 
\frac{\int d\phi\, \chi_2(\phi) e^{-(LT)^3(V_0(\phi) -\chi_2(\phi) \hat\mu_B^2/2)}}{
\int d\phi\, e^{-(LT)^3(V_0(\phi) -\chi_2(\phi) \hat\mu_B^2/2)}}
\end{eqnarray}
This rewriting of the Taylor expansion is a resummation scheme in the sense
that the $n$-th order Taylor coefficient of the pressure ($\chi^{\rm full}_n$) 
is fully accounted
for by the $n$-th order expansion of the effective potential, but the latter
also contributes to all orders higher orders of the ordinary Taylor expansion.
To work out the precise relation we introduce the notation
\begin{eqnarray}
\cavr{X}
&=& 
\frac{1}{Z(\mu_B)}\,\int d\phi\, X(\phi)\, e^{-(LT)^3\,V(\phi)}
\label{eq:doublebracket}
\end{eqnarray}
for an arbitrary integrand $X(\phi)$. One then finds after
inserting Eq.~(\ref{eq:ZwithV}) into the standard definition
\begin{eqnarray}
&&\chi^{\rm full}_{n} =
\frac{1}{(LT)^3}
\left. \frac{\partial^n \,\log\,Z(\mu_B)}{\partial \hat\mu_B^n}\right|_{\mu_B=0}\,,
\label{eq:chifulldef}\\
    \chi_2^{\rm full} &=& \cavr{\chi_2}_0 \,,\label{eq:chi2cavr}\\
    \chi_4^{\rm full} &=& \cavr{\chi_4}_0 + 
    3(LT)^3\left(\cavr{(\chi_2)^2}_0-\cavr{\chi_2}^2_0\right)\,.
    \label{eq:chi4cavr}
\end{eqnarray}
where the $\cavr{\cdot}_0$ integrals are taken at $\mu_B=0$.
The expression multiplied by the volume factor is formally a variance in
the variable $\phi$ and is itself suppressed by volume. Since 
the width of the underlying distribution in $\phi$ is proportional
to $(LT)^{-3/2}$, the term has a finite thermodynamic limit.

\begin{figure*}[t]
\centering
\includegraphics[width=0.99\textwidth]{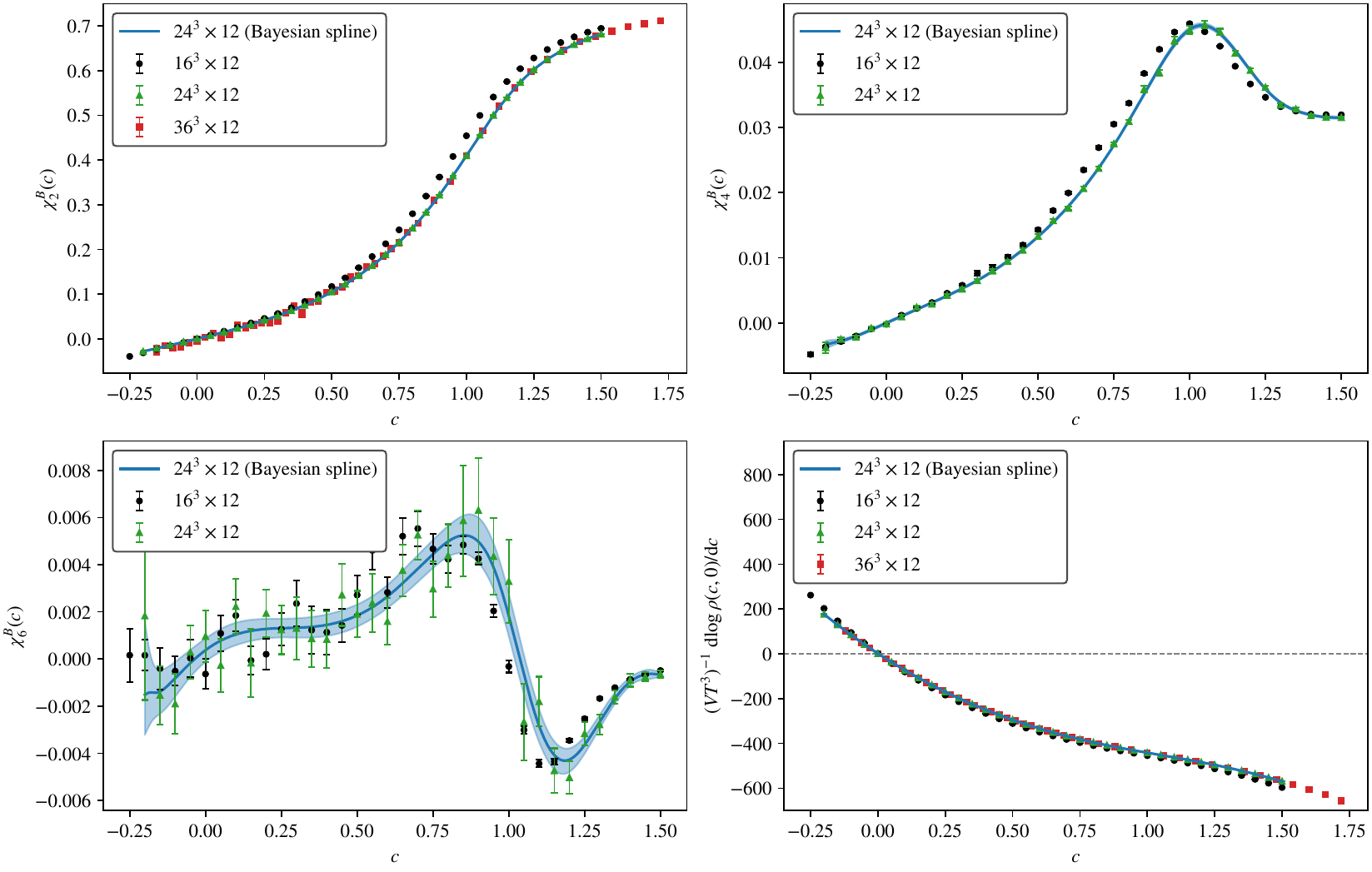}
\caption{\label{fig:coefficients}
The lattice input coefficients for this study at the lowest temperature:
LO ($\chi_2$), NLO ($\chi_4$) and NNLO ($\chi_6$) expansion coefficients of the effective potential. The $\mu_B=0$ potential is found by integrating the bottom right panel's data in $c$. While expectation values have in general severe volume effects in case
of a true transition, the volume dependence of the effective potential is mild.
The band shows the interpolation used for constructing the potential.
}
\end{figure*}

\paragraph{Lattice methods}
We are now explaining how to arrive at the coefficients in 
Eq.~(\ref{eq:Vexpansion}). The standard algorithms sample only the
minimum of the effective potential. With increasing $\mu_B$
the order parameter $\phi$ shifts to a value which would be poorly explored
in the standard setup. Therefore, we compute the entire space of $\phi$
with constrained simulations. We introduce a Gaussian constraint to
the action $\propto -\frac{P}{2}(\phi-c)^2$ where $\phi$ is the real
part of the smeared Polyakov loop and $c$ is the center 
the constraint. This term is implemented in both the molecular dynamics and the
Metropolis steps of the algorithm. We simulate the path integral at fixed $c$
and $\mu_B=0$, the resulting partition sum we call the
\textit{density-of-states} function
$\rho(c)=Z(c,\mu_B=0)$. The constranied ensemble average of $\phi$ gives
\begin{eqnarray}
P \left(\langle \phi  \rangle_c-c\right) = \frac{\partial\log\,\rho(c)}{\partial c}\,
\label{eq:phiandc}
\end{eqnarray}
which can be readily integrated to obtain $\rho(c)$.
For more details of the density-of-states formalism
see Ref.~\cite{Borsanyi:2021gqg}.

With a tight constraint $c$ and $\phi$ can be forced to be arbitrarily close,
and one can consider $c$ as a proxy variable to the actual order parameter
$\phi$. In this sense we determine the potential through
$\log \rho(c) =  -(LT)^3 V(c)$.

The $\mu_B$ dependence of the potential is introduced through the $\mu_B$
derivatives of $\log \rho(c,\mu_B) = \log \,Z(c,\mu_B)$. These are the Taylor
expansion coefficients of the pressure in the constrained ensemble
\begin{eqnarray}
\chi_n(c)=
\frac{1}{(LT)^3}\frac{\partial^n \log\,\rho(c,\mu_B)}{\partial \hat\mu_B^n}
\end{eqnarray}
where $\chi_n(c)$ is computed with the standard stochastic procedure, first
written in Ref.~\cite{Allton:2002zi}, but now applied to each constrained
ensemble. Since the $\mathcal{C}$ symmetry of the constrained ensembles is intact, odd $\chi_n$ coefficients vanish and the even
coefficients are manifestly real, just as in the full ensemble,
as follows from $\gamma_5$-Hermiticity.
The simulation setup is especially simple if an easy-access
gauge observable is chosen as $\phi$, like the Polyakov loop:
this can  be computed independently of the Grassmann integrals.
More precisely, we require an unambiguous definition of $\phi$ for a given gauge configuration. This condition also endorses
the strange condensate as order parameter as long as the chemical
potential is only applied to the light quarks. In contrast,
the chiral condensate would require a different lattice strategy,
since its configuration-specific value directly depends on the light chemical potential. Note that the Polyakov loop couples to the condensate: we observe
suppressed chiral condensates in the presence of larger Polyakov loop constraints.

Technically, the constraint can be applied either by reweighting a high statistics ensemble (e.g. that of Ref.~\cite{Adam:2025phc}) or with dedicated
constrained simulations. We have checked the consistency of these two
possibilities at moderate $\mu_B$. However, the overlap problem inherent
in the finite density reweighting can be evaded only with constrained
simulations, such that the larger values of the order
parameter, typical to high $\mu_B$, are also sampled.

\paragraph{Results}
We present lattice simulations at three temperatures. With a fixed 
lattice spacing ($a=0.2$~fm) we set $T=80$, $96$ and $120$~MeV
in three lattice geometries ($24^3\times12$, $24^3\times10$ and $20^3\times8$,
respectively). Thus, the simulated system size is $L=4 - 4.8$~fm. (See
Refs.~\cite{Borsanyi:2023wno,Adam:2025phc} for simulation parameters and the
details of the 4HEX staggered action.) In each case we performed constrained
simulations at 36 values of the 4HEX smeared bare Polyakov loop.
We computed the generalized quark number susceptibilities $\chi_n(c)$ 
up to sixth order on a statistics of 70k-100k configurations per constraint.
For the lowest temperature we show the resulting coefficients and the derivative
of the density-of-states function at $\mu_B=0$ in Fig.~\ref{fig:coefficients}.
In this plot we also compare three simulation volumes (our main ensemble, a
smaller and a larger lattice) and see mild finite volume effects on the level of
the coefficients. The substantial volume dependence of expectation values and susceptibilities will enter through the explicit volume factor in
Eq.~\eqref{eq:ZwithV}.

\begin{figure*}[t]
\centering
\includegraphics[width=0.99\textwidth]{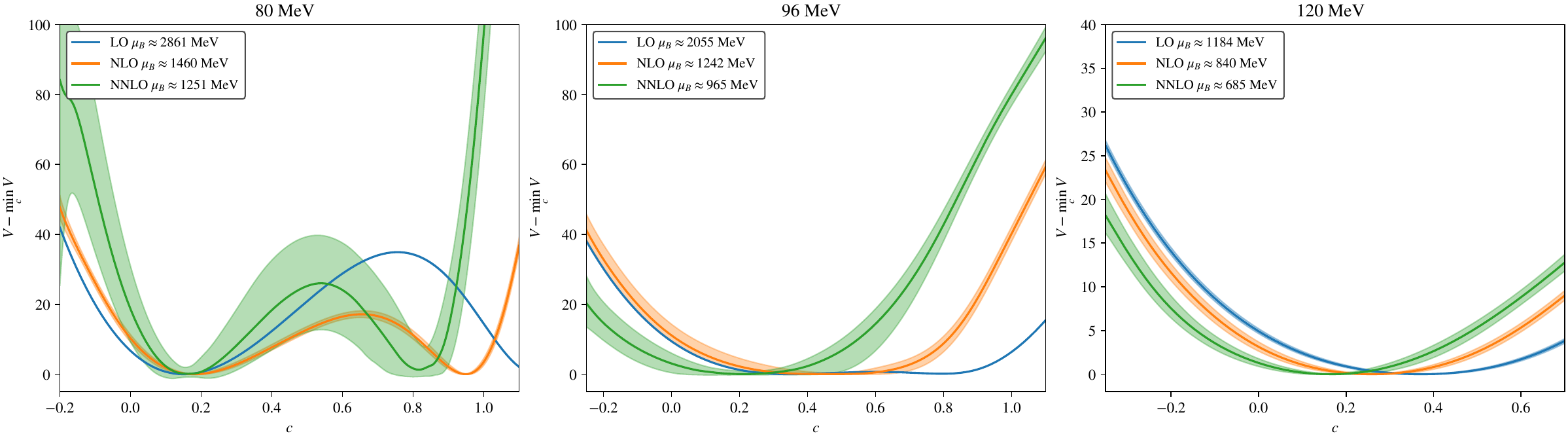}
\caption{\label{fig:potential_order}
Effective potentials at the transition point in $\mu_B$. We show
the potentials at three consecutive orders. Although the actual
transition $\mu_B$ has not converged yet, all three orders give a consistent result on whether at the given temperature a first order, a near-critical or a cross-over transition is expected.
}
\end{figure*}

In Fig.~\ref{fig:potential_scans} we put together the components of
Eq.~(\ref{eq:Vexpansion}). The coefficients are interpolated
first using the Bayesian spline algorithm \cite{Borsanyi:2026ihm}.
This is a Monte-Carlo fitting method of quintic splines that allows the
propagation of the systematic errors of the interpolation to the final result.
The emerging effective potentials are the simple sums of these splines
with some chosen values of the chemical potential. In terms of $\mu_B/T$
these are 15.6, 10 and 5.7 for the three temperatures, respectively. It is
expected that higher orders are necessary before the expansion converges.

Observe, however, that the convergence features of this expansion are
favorable in comparison to the simple Taylor method for observables.
While the coefficients $\chi_n^{\rm full}$ are nearly $n$-independent
at these temperatures \cite{Adam:2025phc}, we see an order of magnitude
drop from $\chi_2(c)$ to $\chi_4(c)$ and a similar factor down from $\chi_4(c)$
to $\chi_6(c)$.

We compare the three orders (LO, NLO, NNLO) in Fig.~\ref{fig:potential_order}.
These differ when plotted at fixed $\mu_B$, but they show a qualitatively
consistent picture when plotted at the respective transition $\mu_B$.
Quite interestingly, the double-well structure at 80~MeV is present at each
of these orders. The flat potential at 96~MeV has two shallow minima at LO
that disappear at high orders. 

\begin{figure*}[t]
\centering
\includegraphics[width=0.99\textwidth]{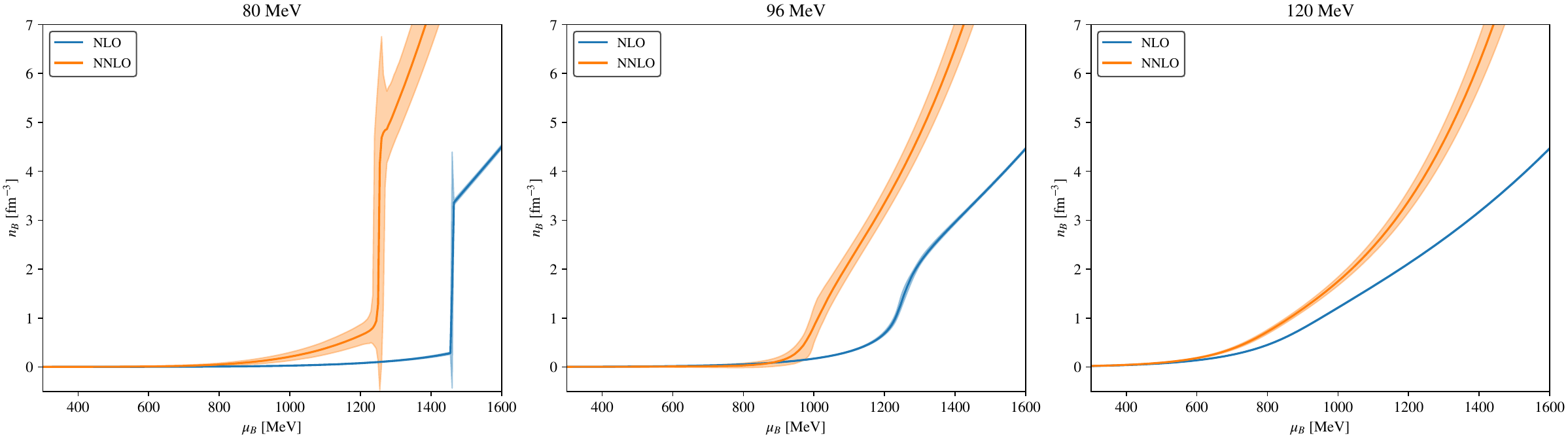}
\caption{\label{fig:nBs}
Expectation value of the baryon number as a function of the chemical potential for the two higher orders of our expansion.
The NNLO contribution outweighs the NLO term, indicating, that
the expansion scheme needs further terms to converge.
}
\end{figure*}

We finally show the expected discontinuity of the baryon density
at the transition in Fig.~\ref{fig:nBs}. The shown functions
are the direct evaluation of Eq.~(\ref{eq:nB}) at the given order.
While at the highest temperature (right panel) this function is continuous,
the discontinuity at the lowest temperature is strongly visible at all studied
orders.

\paragraph{Discussion}
At face value this work could be read as a numerical evidence for a first
order transition in QCD at 80~MeV with hints for criticality near 96~MeV,
thus, demonstrating the existence of a critical end-point.
In fact, several further cross-checks and improvements are necessary before
such a claim can be made.

a) continuum extrapolation: This lattice study was performed at a single
lattice spacing ($a=0.2$~fm) using the 4HEX staggered action with physical
quark masses. We preferred the fixed scale approach to 
the fixed $N_\tau$ approach so that any
differences that we see between temperatures really comes from temperature, and
not from a coupling-dependent bulk transition. This work can be extended to
finer lattices without difficulties other than higher costs. 

b) extrapolation order: We computed coefficients up to $\chi_6$ order. It is
highly desirable to add a further order. While ten-folding the statistics could
make this happen, we must explore the possibility of using additional
imaginary-$\mu_B$ data or to replace the constrained Taylor expansion with an
alternative scheme.

c) simulation volume: Phase transitions are only defined at infinite volume.
However, the extraction of the effective potential is not limited to infinite
volume simulations. Clearly, critical exponents cannot be addressed without
finite volume scaling, but the establishment of a first
order transition is a simpler task, there the volume parameter in
Eq.~(\ref{eq:ZwithV}) is a sufficient tuneable to describe scaling in the
mean-field sense.

We expect that these future improvements will introduce corrections at least
on the quantitative level. 
Should the critical end-point's final position be close to our findings, 
the experimental discovery will face challenges. 
While a position near
$T=96$~MeV is close to other theoretical results \cite{Fu:2019hdw,Gunkel:2021oya},
we find higher chemical potentials placing it significantly away from the chemical
freeze-out curve. If not from freeze-out, one hopes to find evidence
through electromagnetic probes, which then assumes significant time being
spent above and in the transition. Whether a long-lived plasma can be created
in this part of the phase diagram is uncertain.

Another surprising feature of the transition hinted at by our data is the
extreme increase of baryon density. These are higher than what we would see in a
free theory (according to the Stefan-Boltzmann limit). The equation of state
that our
first simulation results suggest would thus disfavor quark matter in
neutron star cores. Mergers, being hotter, may still
sustain a density to form macroscopic
islands of quark matter.

\paragraph{Conclusion} 
We propose a lattice approach to study QCD phase structure at finite density
by shifting the extrapolations in $\mu_B$ from the order parameter's mean
and variance to its effective potential. We introduced constrained simulations
to overcome the strong overlap problem of finite density extrapolations
and worked out the outlined strategy up to NNLO order. We have explicitly 
shown that phase boundaries can be found by this method.

While it is clear that higher orders are necessary, three consecutive orders
presented us a qualitatively consistent picture. These data, which are
not continuum extrapolated, hint for the existence of a critical-endpoint
with a transition temperature close to earlier theoretical expectations, though
at a higher chemical potential. This value of $\mu_B$ has not converged yet,
and shows a decreasing trend as orders are added.

In view of these results we advocate the use density-of-state methods 
not just for
phase structure, but also as a generic resummation device that can boost the
density range of lattice simulation methods.

\paragraph{Acknowledgements}
S.B. thanks for discussions with P. Braun-Munzinger, J. M. Pawlowski and J. Stachel.
This work was supported by the U.S. Department of Energy, Office of Science, Office of Nuclear Physics under contract numbers DE-SC0026065 and DE-SC0022023. Z. Fodor acknowledges funding from the DOE under the contract number DE-SC0025025. This material is based upon work supported by the National Science Foundation under grants No. PHY-2208724, PHY-2116686 and PHY-2514763, and within the framework of the MUSES collaboration, under Grant No. OAC-2103680. This material is also based upon work supported by the National Aeronautics and Space Agency (NASA) under Award Number 80NSSC24K0767. This work has also been supported by the Ministry of Culture and Science of the State of North Rhine-Westphalia, Germany under the funding code NW21-024-A. This work is also supported by the Hungarian National Research, Development and Innovation Office under Project No. FK 147164. A. P. acknowledges support of the Momentum Programme of the Hungarian Academy of Sciences (grant LP2025-13/2025).  This research used resources of the Argonne Leadership Computing Facility, which is a DOE Office of Science User Facility supported under Contract DE-AC02-06CH11357. Access to computer time on Aurora (ALCF) was provided through the INCITE program. The authors gratefully acknowledge the Gauss Centre for Supercomputing e.V. (www.gauss-centre.eu) for funding the this project by providing computing time on the GCS Supercomputer Juwels/Booster at JSC, J\"ulich, Germany. We acknowledge the EuroHPC Joint Undertaking for awarding this project access to the EuroHPC supercomputer LUMI, hosted by CSC (Finland) and the LUMI consortium through a EuroHPC Extreme Access call.

\paragraph{Data availability}
The raw data that support the findings of this article are openly available~\cite{zenodo26doseffpot}.

\bibliographystyle{unsrt}
\bibliography{thermo}

\begin{thebibliography}{10}

\bibitem{Aoki:2006we}
Y.~Aoki, G.~Endrodi, Z.~Fodor, S.D. Katz, and K.K. Szabo.
\newblock {The Order of the quantum chromodynamics transition predicted by the
  standard model of particle physics}.
\newblock {\em Nature}, 443:675--678, 2006.

\bibitem{Borsanyi:2020fev}
Szabolcs Borsanyi, Zoltan Fodor, Jana~N. Guenther, Ruben Kara, Sandor~D. Katz,
  Paolo Parotto, Attila Pasztor, Claudia Ratti, and Kalman~K. Szabo.
\newblock {The QCD crossover at finite chemical potential from lattice
  simulations}.
\newblock {\em Phys. Rev. Lett.}, 125:052001, 2020.

\bibitem{Bazavov:2016uvm}
A.~Bazavov, N.~Brambilla, H.~T. Ding, P.~Petreczky, H.~P. Schadler, A.~Vairo,
  and J.~H. Weber.
\newblock {Polyakov loop in 2+1 flavor QCD from low to high temperatures}.
\newblock {\em Phys. Rev.}, D93(11):114502, 2016.

\bibitem{Borsanyi:2025lim}
Szabolcs Bors\'anyi, Zolt\'an Fodor, Jana~N. Guenther, Ruben Kara, Paolo
  Parotto, Attila P\'asztor, Ludovica Pirelli, and Chik~Him Wong.
\newblock {Chiral versus deconfinement properties of the QCD crossover:
  Differences in the volume and chemical potential dependence from the
  lattice}.
\newblock {\em Phys. Rev. D}, 111(1):014506, 2025.

\bibitem{HotQCD:2018pds}
A.~Bazavov et~al.
\newblock {Chiral crossover in QCD at zero and non-zero chemical potentials}.
\newblock {\em Phys. Lett. B}, 795:15--21, 2019.

\bibitem{Bonati:2018nut}
Claudio Bonati, Massimo D'Elia, Francesco Negro, Francesco Sanfilippo, and
  Kevin Zambello.
\newblock {Curvature of the pseudocritical line in QCD: Taylor expansion
  matches analytic continuation}.
\newblock {\em Phys. Rev.}, D98(5):054510, 2018.

\bibitem{Borsanyi:2024xrx}
Szabolcs Borsanyi, Zoltan Fodor, Jana~N. Guenther, Paolo Parotto, Attila
  Pasztor, Ludovica Pirelli, Kalman~K. Szabo, and Chik~Him Wong.
\newblock {QCD deconfinement transition line up to \ensuremath{\mu}B=400\,\,MeV
  from finite volume lattice simulations}.
\newblock {\em Phys. Rev. D}, 110(11):114507, 2024.

\bibitem{Fu:2019hdw}
Wei-jie Fu, Jan~M. Pawlowski, and Fabian Rennecke.
\newblock {QCD phase structure at finite temperature and density}.
\newblock {\em Phys. Rev. D}, 101(5):054032, 2020.

\bibitem{Gunkel:2021oya}
Pascal~J. Gunkel and Christian~S. Fischer.
\newblock {Locating the critical endpoint of QCD: Mesonic backcoupling
  effects}.
\newblock {\em Phys. Rev. D}, 104(5):054022, 2021.

\bibitem{Wang:2026xwa}
Zi-ning Wang, Li-jun Zhou, Chuang Huang, Rui Wen, Shi Yin, and Wei-jie Fu.
\newblock {Fierz-complete four-quark interactions and the QCD phase diagram}.
\newblock {\em Phys. Rev. D}, 114(9):094028, 2026.

\bibitem{Fu:2026qnl}
Wei-jie Fu, Chuang Huang, Jan~M. Pawlowski, Fabian Rennecke, Rui Wen, and Shi
  Yin.
\newblock {Strangeness neutrality and the QCD phase diagram}.
\newblock 3 2026.

\bibitem{Gao:2026pfr}
Fei Gao, Yuepeng Guan, and Shinya Matsuzaki.
\newblock {Effect of the charm quark on the QCD chiral phase diagram}.
\newblock {\em Phys. Rev. D}, 114(1):014016, 2026.

\bibitem{Allton:2002zi}
C.R. Allton, S.~Ejiri, S.J. Hands, O.~Kaczmarek, F.~Karsch, et~al.
\newblock {The QCD thermal phase transition in the presence of a small chemical
  potential}.
\newblock {\em Phys.Rev.}, D66:074507, 2002.

\bibitem{Fodor:2007vv}
Zoltan Fodor, Sandor~D. Katz, and Christian Schmidt.
\newblock {The Density of states method at non-zero chemical potential}.
\newblock {\em JHEP}, 0703:121, 2007.

\bibitem{Ejiri:2007ga}
Shinji Ejiri.
\newblock {On the existence of the critical point in finite density lattice
  QCD}.
\newblock {\em Phys. Rev. D}, 77:014508, 2008.

\bibitem{Mukherjee:2021tyg}
Swagato Mukherjee, Fabian Rennecke, and Vladimir~V. Skokov.
\newblock {Analytical structure of the equation of state at finite density:
  Resummation versus expansion in a low energy model}.
\newblock {\em Phys. Rev. D}, 105(1):014026, 2022.

\bibitem{Borsanyi:2021gqg}
Szablocs Borsanyi and D\'enes Sexty.
\newblock {Topological susceptibility of pure gauge theory using Density of
  States}.
\newblock {\em Phys. Lett. B}, 815:136148, 2021.

\bibitem{Adam:2025phc}
Alexander Adam, Szabolcs Bors{\'a}nyi, Zoltan Fodor, Jana~N. Guenther, Piyush
  Kumar, Paolo Parotto, Attila P{\'a}sztor, and Chik~Him Wong.
\newblock {High-precision baryon number cumulants from lattice QCD in a finite
  box: Cumulant ratios, Lee-Yang zeros, and critical endpoint predictions}.
\newblock {\em Phys. Rev. D}, 113(7):074525, 2026.

\bibitem{Borsanyi:2023wno}
Szabolcs Borsanyi, Zoltan Fodor, Jana~N. Guenther, Sandor~D. Katz, Paolo
  Parotto, Attila Pasztor, David Pesznyak, Kalman~K. Szabo, and Chik~Him Wong.
\newblock {Continuum-extrapolated high-order baryon fluctuations}.
\newblock {\em Phys. Rev. D}, 110(1):L011501, 2024.

\bibitem{Borsanyi:2026ihm}
Szabolcs Bors{\'a}nyi, Zolt{\'a}n Fodor, Jana~N. Guenther, Piyush Kumar, Paolo
  Parotto, Attila P{\'a}sztor, and Chik~Him Wong.
\newblock {Expansion schemes for finite density QCD equation of state: which
  one is the best?}
\newblock 10 2026.

\bibitem{zenodo26doseffpot}
Wuppertal-Budapest collaboration.
\newblock Data set for hints for a qcd critical end-point from simulations on a
  finite lattice.
\newblock 10.5281/zenodo.24595853.

\end{thebibliography}

\end{document}